# FUNHOUSE MIRROR OR ECHO CHAMBER? A METHODOLOGICAL APPROACH TO TEACHING CRITICAL AI LITERACY THROUGH METAPHORS

Jasper Roe [1*], Leon Furze [2], Mike Perkins [3]

[1] James Cook University Singapore, Singapore.
[2] Deakin University, Australia
[3] British University Vietnam, Vietnam.

[*] Corresponding Author: jasper.roe@jcu.edu.au

November 2024

## Abstract

As educational institutions grapple with teaching students about increasingly complex Artificial Intelligence (AI) systems, finding effective methods for explaining these technologies and their societal implications remains a major challenge. This study proposes a methodological approach combining Conceptual Metaphor Theory (CMT) with UNESCO's AI competency framework to develop Critical AI Literacy (CAIL). Through a systematic analysis of metaphors commonly used to describe AI systems, we develop criteria for selecting pedagogically appropriate metaphors and demonstrate their alignment with established AI literacy competencies, as well as UNESCO's AI competency framework.

Our method identifies and suggests four key metaphors for teaching CAIL. This includes GenAI as an echo chamber, GenAI as a funhouse mirror, GenAI as a black box magician, and GenAI as a map. Each of these seeks to address specific aspects of understanding characteristics of AI, from filter bubbles to algorithmic opacity. We present these metaphors alongside interactive activities designed to engage students in experiential learning of AI concepts. In doing so, we offer educators a structured approach to teaching CAIL that bridges technical understanding with societal implications. This work contributes to the growing field of AI education by demonstrating how carefully selected metaphors can make complex technological concepts more accessible while promoting critical engagement with AI systems.

**Keywords**: Critical AI literacy, artificial intelligence, metaphors in education, GenAI,

# Introduction

The quick emergence and broad adoption of Generative Artificial Intelligence (GenAI) have created an urgent imperative for educational institutions to adapt their approaches to teaching and learning. As these technologies become increasingly embedded in academic and professional contexts, educational researchers are grappling with fundamental questions about how to effectively use, teach, and critically engage with GenAI in schools and universities. This shift has catalysed the development of a new field of study focused on AI literacy, with governments and educational bodies worldwide recognising the vital importance of preparing learners at all educational levels for an AI-enabled future (Miao & Kelly, 2024), including at the earliest levels of education (Su et al., 2023). AI literacy has already been introduced in various national curricula worldwide (Laupichler et al., 2022; Sperling et al., 2024); however, AI literacy is a new, loosely defined, and inconsistently applied concept, with no universally agreed upon definition (Bozkurt et al., 2023). Nevertheless, AI literacy is of central importance in education, as students will inevitably need to engage with AI in various ways in the future, as one aspect of broader technological multiliteracy (Stolpe & Hallström, 2024).

In this study, we adopt the AI literacy definition proposed by Long and Magerko (2020), who define it as a set of competencies which enable users to critically evaluate AI technologies, use them effectively for collaboration and communication, and do so in multiple contexts, including at home, work, and online. In one of the few publications on AI literacy that specifically focuses on and distinguishes GenAI, that is, AI applications which can produce multimodal outputs, Bozkurt et al. (2023) argue that GenAI literacy must go beyond mere basic understanding, requiring a comprehensive approach that integrates theoretical knowledge, practical skills, and deep critical reflection, presenting a framework based on three areas: Know What (theoretical knowledge), Know How (practical application), and Know Why (ethical and philosophical understanding) which aligns with some of the areas proposed in other AI competency frameworks (Chiu et al., 2024). By promoting GenAI literacy, learners will be better placed to harness these tools and ensure that they are used to maximise benefits and minimise negative impacts (Bozkurt et al., 2023).

However, we extend this concept further to what we conceptualise as Critical AI Literacy (CAIL). We define CAIL as:

The ability to critically analyse and engage with AI systems by understanding their technical foundations, societal implications, and embedded power structures, while recognising their limitations, biases, and broader social, environmental, and economic impacts. This literacy enables both practical engagement with AI and a critical reflection on its ethical implications.

Critical AI Literacy can be seen as stemming from related fields such as Critical Digital Literacy (CDL) and Critical Literacy. CDL, in its simplest form, relates to the critical consumption of digital media (Pangrazio, 2016); however, in practice, it extends to a deeper critique of the architecture and systemic structures of digital ecosystems (Knight et al., 2020). This approach, drawing from Ávila and Pandya (2012) and Jenkins (2006), empowers individuals as both consumers and creators of digital content. It examines how "digital (inter)actions" (Knight et al., 2020, p. 20) influence our engagement with technology, emphasising the need to scrutinise nonhuman agents, such as algorithms and chatbots, which often shape user experiences.

Central to CDL is the interrogation of "platform ecologies" (Garcia & Nichols, 2021), which links user behaviours with the underlying design and resource requirements of digital platforms (Van Dijck, 2021). This perspective critiques the agency of digital platforms by revealing the power structures embedded within their interfaces and algorithms and positioning CDL as a tool for understanding the influence of design on user interaction. By investigating these architectures, CDL highlights the importance of considering how digital systems reinforce or disrupt existing social and power dynamics, aligning with broader sociocultural critiques (Emilia Djonov & van Leeuwen, 2017).

Drawing on these approaches, CAIL involves critically analysing AI's design, biases, and implications, reflecting on how metaphor and discourse shape our understanding and use of AI. Bali (2023) highlights that CAIL includes questioning the assumptions of AI's intelligence and examining its role within broader societal and ethical contexts. By understanding AI through critical frameworks, educators and students can develop a nuanced perspective that transcends instrumental use and encourages a reflective, socially aware approach to AI adoption and use in educational contexts (Gupta et al., 2024).

Through CAIL, we posit that critically evaluating AI technologies is not sufficient; rather, we wish to foster an understanding and deep reflection among all learners of the potentially disastrous consequences of unregulated AI (OECD, 2024); the extreme social, environmental, and economic costs of AI development (Driessens & Pischetola, 2024); and the cultural biases that typify some GenAI models (Roe, 2024). This includes facing the complex reality that, while we may still use and benefit from AI tools, we must not turn away from the questionable, alarming, and at times frightening aspects of new technologies.

**Conceptual Metaphor Theory**

Teaching CAIL presents unique pedagogical challenges, as educators must help learners grasp the complex technological systems and their societal implications in a nuanced and objective way. Metaphors may offer a valuable solution by serving as a powerful learning resource for sparking discussions and creative thinking about AI systems. Although the concept of metaphor is familiar to most people through formal schooling, informal conversation, and popular media, its scholarly study has evolved into a rich field of enquiry. The study of metaphor is embedded in multiple frameworks that aim to describe topics such as language, thought, and communication (Gibbs Jr., 2008), with multiple handbooks, volumes, and academic journals dedicated to investigating the relationship between metaphors and the social world. This theoretical foundation suggests that metaphorical approaches are particularly useful for developing CAIL.

In this paper, we frame our understanding of metaphor through Conceptual Metaphor Theory (CMT), developed in Lakoff and Johnson's seminal book *Metaphors We Live By* (Lakoff & Johnson, 2003). Following this theoretical framing, we contend that metaphors are not simply turns of phrase or rhetorical flourishes, but are powerful, pervasive conceptual tools for structuring, restructuring, and even creating reality (Batten, 2012). In essence, this form of metaphor is operationalised through understanding and experiencing one kind of thing in terms of another (Lakoff & Johnson, 2003). Furthermore, metaphors can fulfil multiple social and cognitive functions, ranging from strategic persuasive devices (Ferreira et al., 2023) to act as constructs to help us organise our knowledge of the world (Saban, 2006) and understand abstract concepts (Niemeier, 2017). In explaining the relationship between metaphor and thought, Kövecses (2020) provides an example of the 'journey of life'. Such usage of the 'journey' metaphor does not necessarily only refer to how we speak about life, but also how we think about life, and subsequently, how we may go on to act in this belief that life is a journey. Given that fostering CAIL requires critical thinking and that metaphors can help us restructure our reality (Batten, 2012), there is a natural alignment between these two concepts for pedagogical use.

**Metaphor in Education**

Historically, metaphors have been used to facilitate education to expand students' minds and promote critical thinking (Low, 2008). Instruction often requires the description and discussion of complex meanings to progress understanding, and metaphors are appropriate tools for achieving this objective (Carter & Pitcher, 2010). As instruction requires moving from known to unknown and concrete to abstract, metaphors can use concrete examples to explain abstract principles (Clarken, 1997) and can be used for a variety of learning activities, such as finding memorable labels for complex concepts or helping learners understand challenging learning materials (Low, 2008). Metaphors may be employed as a powerful learning mechanism even in early childhood, with studies showing that three- and four-year-olds can use metaphors to make inferences about the functional features of objects (Zhu & Gopnik, 2023). Metaphors have also been studied in other disciplines which require close social relations and dialogue, such as psychotherapy, with authors noting that the use of appropriate metaphor can lead to positive client outcomes and cognitive engagement (McMullen & Tay, 2023). Indeed, there is a rich body of literature describing how metaphors are used to refer to the teaching and learning process itself (Alger, 2009; Batten, 2012, 2012; Clarken, 1997; Hager, 2008; Saban, 2006), with Hager (2008) claiming that humans are unable to think about learning without employing some form of metaphor.

Recent research has empirically validated the value of metaphors in various classroom contexts in higher education. For example, Pager-McClymont & Papathanasiou (2023) demonstrated the value of using CMT to teach English for Academic Purposes (EAP), by using the 'A is B' metaphor structure to teach composition, describing arguments (A) as buildings (B) and writing (A) as cooking, eating, and digesting (B). In another related example, Haidet et al. (2017) explored the use of jazz music as a metaphor for teaching effective communication strategies in patient-doctor interaction, noting positive results in using this metaphor to engage students engaged in medical training.

At the same time, it is important to be aware of the limitations of metaphor use in an educational environment. Inappropriate uses, opaque uses, or metaphors which do not have a high degree of similarity between the two topics may create difficulties for learners or impede the educational process. The question of culture also arises when selecting metaphors, especially when teaching a foreign language. To illustrate, Low (2008) notes that as metaphors may be culturally specific, if teaching learners the skill of language proficiency in a foreign language, there is an argument as to whether it is more appropriate to teach cultural concepts first rather than teach the metaphor. Furthermore, the limits of metaphors are an important consideration when teachers are engaged in the process of selecting one to use. Carter and Pitcher (2010) explain this by drawing attention to the ubiquitous use of metaphors in teaching electricity. The authors explain that when teaching the concept of electricity, the metaphor of electricity as water is commonly applied. The reasoning behind this is that there is a perceived similarity in the 'flow' between electrons in a wire and water in a pipe. However, they highlight that extending this metaphor results in a conceptual break, as there are significant differences in some areas of flow; electron flow cannot be seen, whereas water can, and a broken pipe results in water continuing to flow, but electricity in a wire will stop (Carter & Pitcher, 2010). Consequently, while there is significant evidence that metaphors are appropriate and beneficial concepts to use in educational processes, their usefulness is limited; metaphors must be appropriate and understandable, and the limits of comparison should be considered by the educator. We draw on these principles to develop a set of criteria that can be applied when choosing to use metaphors to illustrate information about AI technologies.

**Metaphors and Artificial Intelligence**

AI is no stranger to metaphors, and in the current era of developing AI technologies, metaphors are increasingly common, although some have been described as problematic and anthropomorphising (Furze, 2024). At the same time, despite the burgeoning research on AI and GenAI, our initial literature search using the Scopus and Web of Science (WoS) databases resulted in few recent studies on the relationship between AI and metaphor. That said, despite not being a highly active area of research, explorations of metaphors and AI have been taking place for a significant period. Gozzi (1994), for example, used a dataset of articles in the press at the earliest stages of development in the field of AI, from 1966 to 1970, and found that common metaphors for computers were 'brains' and that this also extended to the description of thought as a computational process, resonating with the common metaphors for AI that we encounter in popular media today.

More recent work on the intersection of AI and metaphor has explored the use of AI in science fiction as a metaphor for other aspects of human existence (Hermann, 2023), and studies have been undertaken to ascertain whether LLMs, such as GPT-4, can interpret literary metaphors (Ichien et al., 2024). Carbonell et al. (2016) describe the connections between computational metaphors for the brain and vice versa, arguing that this reinforces the relationship between the brain and computers, and that this subsequently has an effect that shapes society. In the realm of explainable AI, ways of explaining AI chatbots using the metaphor of fermentation and bread-making have also been explored (Nicenboim et al., 2023). Nguyen (2024) conducted a semiotic analysis of ChatGPT's self-representation, noting that the language model seemed to imply a metaphorical representation of itself, and Hunger (2023) describes the historical process of anthropomorphising metaphors related to AI.

Rehak (2021) argues that powerful metaphors of AI are used to perpetuate the deterministic myths of AI technology. Even the term Artificial Intelligence itself is problematic; much of the technology is neither artificial nor intelligent (Crawford, 2021). Machine learning algorithms and datasets which underpin these systems also serve to distance the output from human responsibility, complicating earlier ethical debates (Campolo & Crawford, 2020).

The language surrounding the processes of AI are also commonly anthropomorphised. Rehak (2021) points to words such as "'recognition', 'learning', 'acting', 'deciding', 'remembering', 'understanding'" in processes carried out by machine learning algorithms, comparing the anthropomorphic words used in the field of AI to the more abstract language typically used in mathematics. The Royal Society (2018) are critical of this kind of language as part of the AI narrative, claiming that it causes a disconnect from the technology itself which may contribute to a "hype bubble", public fears about technology and subsequent reluctance to adopt beneficial technologies, and a distortion of the discourse surrounding the future of digital technology.

One recent work by Anderson (2023) focuses on the metaphorical framing of Large Language Models (LLMs) such as ChatGPT as either a tool or a collaborator, calling for activities which enable us to understand the biases, inaccuracies, and faults of such tools and foster the development of students' digital literacy. At the same time, the author points out that there is a need to study how metaphorical language is used when discussing ChatGPT, as such uses could complicate, rather than aid, how we understand these technologies. Furthermore, conveying the functions and malfunctions of LLMs using precise and accurate metaphors can help foster an understanding of how these tools work among the public and academics (Smith et al., 2023). Ye & Li (2024) used a corpus linguistics approach to investigate conceptual metaphor in the European Union AI Act (EU AIA), finding that metaphors used included those related to the concepts of 'Journey', 'Human', 'War', and 'Object'.

Specific research related to education has explored how learners in educational contexts perceive AI technologies through metaphors. Yan et al. (2024) investigated Chinese EFL learners' metaphorical conceptualisations of GenAI and found that participants viewed GenAI through multiple metaphorical categories such as humans, tools or machines, brains, resources, food and drink, and medicine. Given this diversity of conceptualisation, the authors call for further work to promote digital literacy for students using GenAI. A similar study was undertaken with student teachers in Germany (N=100) (Şentürk & Akol Göktaş, 2024), which found that participants conceptualised GenAI as a library, student, talking pool, warehouse, world, star, game, human, and brain, among others. In relation to Yan's (2024) findings, corresponding findings of categories such as 'humans' offer insight into potential cross-cultural similarities in how we use metaphor to structure understanding of these technologies.

Gupta et al. (2024) explored how using discussions related to metaphors for AI can assist in developing awareness of how AI systems work, as a method of fostering Critical AI Literacy. By engaging in a digital collaborative autoethnographic study, the authors developed a set of metaphors derived from popular online media, scholarly literature, social media, and research participants. The authors contend that such an approach is pedagogically powerful as it can stimulate the affective domain, thus benefiting learning. Crucially, these authors offer advice for teaching CAIL through metaphors, which we foreground in our work, suggesting that educators could ask students to share metaphors that they have discovered, and then challenge them to come up with additional input to extend or offer a new metaphor to deepen their understanding of these complex concepts.

## Methodology

**Framework Development Process**

To demonstrate the ways in which CAIL can be fostered by using metaphor, we opted to develop a criteria and rating scale to choose appropriate metaphors and then create sample learning activities which are linked to the UNESCO AI competency framework for students (Miao & Kelly, 2024). Our first task was to generate a list of popular metaphors used to describe AI systems to have widespread appeal and familiarity with both teachers and students. To do this, we adopted a similar approach to Gupta et al. (2024), drawing on existing literature on traditional and online media, social media, and scholarly work to arrive at a list of commonly used metaphors. We then developed a set of criteria against which to measure the suitability of these metaphors for classroom activities and assessed their ability to foster AI literacy by evaluating their alignment with curricular goals for developing AI literacy in UNESCO's AI competency framework for students (Miao & Kelly, 2024).

The UNESCO AI competency framework for students outlines three levels of competency: Understanding, Applying, and Creating. Among these, the most suitable area for CAIL to be developed using metaphors is understanding. Miao and Kelly (2024) explain that at this level of understanding, students are required to develop an awareness of what AI is and be able to interpret different ethical issues, technical knowledge, and underlying processes that power AI. The authors also suggest that real-life practices should be integrated to help support understanding. Following these principles and through discussion and consensus building, we created four criteria for evaluating the appropriateness of metaphors to guide the understanding of these concepts: Accessibility, Explanatory Power, Critical AI Literacy Potential, and Pedagogical Utility. Descriptions of these criteria are presented in Table 1.

*Table 1: Criteria for Evaluating Appropriacy of AI Metaphors*

| Criteria for Metaphor Evaluation | Rationale |
|---|---|
| Accessibility | An appropriate metaphor will be familiar to students and accessible to a general audience. |
| Explanatory Power | An appropriate metaphor has the potential to illuminate key aspects of GenAI systems and develop understanding. |
| Critical AI Literacy Potential | An appropriate metaphor may encourage critique of AI systems limits and capabilities. |
| Pedagogical Utility | An appropriate metaphor will support the creation of varied learning activities. |

After deciding on the criteria for appropriate metaphor selection, we drew up a list of ten common AI metaphors that we located in previous scholarly literature and popular online media. The ten selected metaphors are listed in Table 2.

*Table 2: Initial List of Potential Metaphors to Guide AI Literacy*

| Description | Description |
|---|---|
| Stochastic Parrot (Bender et al., 2021) | Coined in a highly cited paper to refer to the probabilistic, non-understanding nature of AI language models. |
| Black Box (Furze, 2024) | An opaque system with observable input and outputs but no access to inner processes. |
| Iceberg (Furze, 2024) | Generative AI is powered by an enormous, largely unknowable dataset 'below the waterline'. Consumers and users only interact with a small portion of the model. |
| Funhouse Mirror | Provides a reflected version of reality but in distorted and warped ways. |
| Assistant | An aide that can assist with simple tasks but requires supervision. |
| Loudspeaker (Gupta et al., 2024) | Amplifies and broadcasts existing patterns. |
| Double Edged Sword | A tool or weapon with both beneficial and harmful aspects. |
| Calculator for words (Willison, 2023) | A mathematical calculation of language. |
| Natural disaster | A powerful force that can be prepared for, but not avoided. |
| Collaborative Artist | A creative partner that can contribute to an artistic process. |
| Map | A representative map of society and culture based on the training data. |
| Pattern matching machine (Furze, 2024) | A system that identifies and reproduced patterns from data. |
| Echo Chamber | A system that reflects and reinforces existing patterns. |

The selection of metaphors for deeper exploration followed a qualitative, consensus-building approach guided by both established principles from metaphor research and the UNESCO AI competency framework (Miao & Kelly, 2024). Our process was fundamentally grounded in Lakoff and Johnson's (2003) Conceptual Metaphor Theory, which establishes metaphors as tools for structuring thought and understanding reality and supports mapping between metaphor source domains and the target domain (AI).

Through collaborative discussions, our research team jointly evaluated each metaphor against our established criteria (Accessibility, Explanatory Power, Critical AI Literacy Potential, and Pedagogical Utility) and their alignment with UNESCO's AI competency goals. This qualitative analysis was mindful of Low's (2008) emphasis on cultural accessibility, and Carter and Pitcher's (2010) caution about how metaphors can break down when extended too far. Batten's (2012) work on how metaphors can reconstruct understanding in educational contexts guided our analysis of how each metaphor functions pedagogically. Recent empirical studies by Yan et al. (2024) and Şentürk and Akol Göktaş (2024) validated our approach, demonstrating that learners naturally conceptualise AI through multiple metaphorical categories; therefore, we sought broadness across the chosen metaphors, making a final selection based on a combination of alignment with our selected criteria, as well as the UNESCO AI competency curriculum goals and elements of understanding in the AI competency framework that they most closely matched (Miao & Kelly, 2024). We put these into an 'A is B' structure, as shown in Table 3.

*Table 3: Selected Metaphors, Selection Criteria and Alignment to UNESCO AI Competency Framework*

| AI Metaphor | Relation to selection criteria | Alignment to UNESCO AI Curriculum Goals |
|---|---|---|
| AI is a Funhouse Mirror | **Accessibility:** The concept of a funhouse mirror is familiar across age groups<br>**Explanatory Power:** Illustrates how AI systems may distort reality<br>**Critical AI Literacy Potential:** May lead to discussion about bias and representation<br>**Pedagogical Utility:** Lends itself to physical activities with realia (e.g. a distorted mirror) and links to activities exploring data and algorithmic bias | CG4.1.3.2 Develop conceptual knowledge on how AI is trained based on data and algorithms |

| AI Metaphor | Relation to selection criteria | Alignment to UNESCO AI Curriculum Goals |
|---|---|---|
| AI is a Map | **Accessibility:** Draws on familiar concepts of maps as a way of representing the world<br>**Explanatory Power:** Demonstrates how AI is a representation but not a true reflection of the world<br>**Critical AI Literacy Potential:** Encourages the examination of power structures and colonialism<br>**Pedagogical Utility:** Allows for debate and critical thinking regarding power and representation | CG2.1.1 Surface ethical controversies through a critical examination of use cases of AI tools in education. |
| AI is an Echo Chamber | **Accessibility:** Echoes a universal physical reality across cultures<br>**Explanatory Power:** Demonstrates how AI systems may reinforce ideas, biases, or concepts in the training data<br>**Critical AI Literacy Potential:** Leads to the critical discussion of how to mitigate feedback loops and filter bubbles<br>**Pedagogical Utility:** Supports numerous practical and personal activities, for example exploring algorithmic advert selection on social media | CG4.1.4.1 Scaffold critical thinking skills on when AI should not be used |
| AI is a Black Box | **Accessibility:** The concept of a black box is a long-standing metaphor with high familiarity<br>**Explanatory Power:** Helps illustrate the challenge of understanding AI systems<br>**Critical AI Literacy Potential:** May promote discussion of transparency and explainability<br>**Pedagogical Utility:** May be demonstrable by encouraging learners to generate unexplainable outputs | CG4.1.2.1 Illustrate dilemmas around AI and identify the main reasons behind ethical conflicts |

## Teaching Activities for AI Literacy

To illustrate our conceptual method for teaching AI literacy by exploring conceptual metaphors, we created sample learning activities for each of the four metaphors. We began by defining our learning outcomes based on UNESCO AI curriculum goals contained in the competency framework (Holmes & Kelly, 2024), and then drawing on our professional knowledge as educators to devise appropriate exercises for a multidisciplinary higher education classroom. The activities were structured by first introducing the metaphor in question, followed by a scaffolding learning activity, and then a final teacher-facilitated discussion.

**Activity 1: AI is a Funhouse Mirror**

*Learning Objectives*

1. Recognize how AI systems may distort information
2. Understand the principles of bias in AI outputs
3. Explore the appropriacy of a Funhouse Mirror metaphor in relation to AI systems

*Addresses Curriculum Goal*

CG4.1.3.2 Develop conceptual knowledge on how AI is trained based on data and algorithms.

*Introducing the metaphor*
The instructor can begin by showing an image of a funhouse mirror and eliciting experiences from the learners; for example, if they had encountered a funhouse mirror before, the intended or unintended effects of the mirror, and its form and purpose. The instructor can then explain the relationship between AI systems and a funhouse mirror, namely, that they may produce distorted versions of reality.

*Learning Activity: Prompt testing to observe bias*
In small groups of three to four students, have learners search for articles on bias and distortions in GenAI outputs and imagery using the Internet. Following this, learners should be encouraged to test the same prompt in different GenAI systems and document any variations in their responses. Have learners compare the outputs and identify the differences, and then ask learners to discuss whether these could represent 'reflections' of potential biases in training data.

*Discussion Questions*
The instructor may prompt learners to consider whether there are patterns of distortion that are consistent across systems and identify the potential implications of these biases. Finally, ask students to reflect on how accurate the funhouse mirror metaphor is, prompting them to consider what the impacts of distorted GenAI outputs could be on the individual or society.

**Activity 2: AI is an Echo Chamber**

*Learning Objectives*

1. Understand the concept of a feedback loop
2. Explore the concept of a 'filter bubble'
3. Explore the appropriacy of an Echo Chamber metaphor in relation to AI systems

*Addresses Curriculum Goal*

CG4.1.3.2 Develop conceptual knowledge on how AI is trained based on data and algorithms.

*Introducing the Metaphor*

The instructor can begin by introducing the echo chamber concept. Multiple forms of media can be used to demonstrate this, for example, a cartoon, video of an echo chamber, or audio clips of an echo. The instructor can ask learners to reflect on why an AI system can be seen as an echo chamber.

*Learning Activity: Observe the echo-chamber effect.*

In this activity, learners can actively explore how limited training data and repeated prompting of a GenAI tool can lead to a narrowing set of options. Encourage students to experiment with multiple GenAI tools with a neutral prompt, for example, asking for a recipe for a healthy breakfast. Then, learners should be encouraged to continue to prompt for further examples and examine the way in which the opinions and responses become increasingly narrow. As a follow-on activity, learners can be asked to reflect on the types of advertisements that they see online and compare them to peers, then reflect on the nature of algorithmic suggestions as a way of creating 'filter bubbles'.

*Discussion Questions*

Following this activity, the instructor may ask learners to discuss the effects of filter bubbles and their potential effects (e.g. polarisation), as well as how algorithmic systems can produce undesired or unintentional consequences. This is used as a basis for discussing whether AI corporations have an ethical responsibility to reduce personalisation through algorithmic means and the pros and cons of this approach. Finally, the instructor may ask learners to reflect on whether the 'echo chamber' metaphor is an appropriate way of describing AI systems.

**Activity 3: AI as a Map – Representation, Power and Bias**

*Learning Objectives*

1. Analyse how AI's representation of knowledge parallels historical maps, focusing on inclusivity, bias, and power.
2. Recognise limitations in AI's representation of knowledge.
3. Critique the metaphor of AI as a map to uncover insights into technology's impact on perception and inclusivity.

*Addresses Curriculum Goal*

CG2.1.1: Surface ethical controversies through a critical examination of the use cases of AI tools in education.

*Introducing the Metaphor*

The instructor can begin by discussing historical maps, such as the Mercator projection, which often distorts size and centralises Western countries, as a springboard to AI's role in shaping perspectives. Students could explore questions such as what parts of reality are emphasised or minimised in a map, and why? Who decides what is "on the map," and how does this affect our understanding of the world?

The instructor can introduce AI as a map, illustrating that AI "maps" knowledge through data selection, categorisation, and emphasis, with similar power dynamics shaping what is visible or hidden. Reflect on the statement "the map is not the territory," exploring how the representation of the world based on the Large Language Model dataset is not a true reflection

*Learning Activity: Mapping AI's Knowledge Terrain*

In small groups, students will choose an area of knowledge (e.g. cultural heritage, health information, and social trends) and map it from the perspective of an AI tool. They should analyse how AI represents this area, noting:

- What information is readily accessible online that can be "mapped" by data scraping?
- What perspectives or voices are missing or minimised from the data?
- What biases or assumptions are present in AI output? (in terms of the metaphor, how is the "map" different to the "territory" or reality?)

Compare and Contrast "Maps": Groups can then compare their findings by examining differences in representation and potential biases.

*Discussion Questions*

In analysing AI's 'map' of knowledge, students observe which perspectives are amplified and which are neglected, noting that AI often prioritises dominant narratives while overlooking marginalised voices. This selective representation shapes our perception of knowledge, as AI's outputs reflect the biases and gaps inherent in its training data. The metaphor of 'AI as a Map' highlights AI's impact on knowledge and power by revealing how certain viewpoints are centred while others are diminished, much like historical maps that emphasise the perspectives of those in control. However, this metaphor has limitations, as it may imply a static view of knowledge rather than AI's dynamic interaction with evolving data. Understanding these dynamics encourages a more responsible approach to AI use, prompting users to critically assess an AI's outputs and recognise where broader, more inclusive perspectives are needed.

**Activity 4: AI is a Black Box**

*Learning Objectives*

1. Develop understanding of the current limitations of AI transparency
2. Recognize the importance of verification of AI generated information
3. Explore the appropriacy of the AI as a Black Box metaphor

*Introducing the metaphor*

Show learners an example of a 'black box' either physically or as an image. Using GenAI imagery to create a picture of a black box can also be an interesting example to demonstrate the metaphor in question and can be a departure point of analysis (for more examples of analysing bias in GenAI imagery, see Roe (2024)).

*Learning Activity: Exploring randomness in GenAI output*

Encourage learners to explore randomness in the outputs of GenAI tools through experimentation. For example, ask learners to prompt multiple GenAI tools to finish a simple sentence, such as 'the cat sat on the…' and document the different responses. This demonstrates that GenAI tools produce unexplainable random outputs that do not necessarily conform to an expected logical output. Learners can then compare their output to one another. Using imagery or video can also be an effective option, using freely available tools such as Windows Image Generator, although this requires signing up for a free account. Attempting to interpret and analyse GenAI images can help to identify why the output is unexplainable.

*Discussion Question*

Ask learners to engage in a discussion to consider why explainable AI matters, the implications of randomness in certain situations, and the potential consequences; for example, if such models should be deployed in high-stakes environments (e.g. making medical or legal decisions, writing important documents) or in lower-stakes environments such as generating code or creating content. Following this, have students reflect on whether the black box metaphor is appropriate for describing the outputs of GenAI tools and AI systems more generally. This can be a time to reflect on the importance of AI literacy and the effects if users are unaware of the randomness inherent in some AI system outputs.

## Limitations

Our proposed methods for teaching AI literacy through metaphors offer an additional lens for fostering an understanding of how AI systems work in an accessible, interesting, and engaging manner. At the same time, multiple limitations should be considered when implementing this approach in classrooms. First, consideration of metaphor appropriacy is key, as metaphors can break down when extended too far away (Carter & Pitcher, 2010). This is relevant when discussing AI systems, as metaphors may be effective in some respects. For example, the funhouse mirror effectively illustrates the concept of distortions in GenAI outputs; however, it does not adequately capture the algorithmic processes that produce these distortions. For this reason, these metaphors should be viewed as objects of critique and a starting point for further discussion.

Metaphor choice should be considered from a culturally diverse perspective as metaphors may be culturally specific. Therefore, it is important to choose metaphors that are accessible to target learners. The 'funhouse mirror' metaphor specifically, may not be familiar to learners from a culture that does not have access to these attractions. Similarly, the concept of a map can vary across cultures. This is a limitation of the approach, but it is not insurmountable with careful planning.

Finally, although we attempted to adopt a systematic and methodologically rigorous approach to selecting and evaluating metaphors, we relied on qualitative assessment and consensus building among researchers to validate the effectiveness of these metaphors in promoting AI literacy. Finally, given the recent nature of calls to promote AI literacy, with the UNESCO AI competencies framework released only in 2024, practical validation of these exercises has not yet been carried out.

## Conclusion

As AI technologies continue to develop at a breakneck speed and become more embedded in daily life, the need for CAIL in education continues to increase. Multiple approaches from different disciplinary perspectives are needed to help learners engage in these systemic changes in society. This paper argues that the growing discourse surrounding the appropriacy of metaphors for AI in society can be used as a helpful learning resource for fostering AI literacy, building on the long history of using metaphors in educational contexts. Our development of criteria for assessing metaphors for use in this context is intended to be simple, practical, and usable for educators, while our selection of four metaphors that describe unique attributes of AI systems exemplifies how this approach can be enacted in the classroom and how it can be aligned with an existing AI competency framework. The four key metaphors we describe — the funhouse mirror, echo chamber, map, and black box—provide educators with examples of potential learning activities.

The value of this CMT-driven approach to teaching AI literacy enables educators to address multiple aspects of AI literacy simultaneously, as learners can engage with technical concepts and the current literature while equally examining the ethical, societal, and individual impacts of AI. At the same time, we highlight that this theoretical paper should serve as a starting point for future research, including longitudinal studies which examine the efficacy of metaphor-based AI literacy education, the refinement and validation of a set of selection criteria, investigation of how metaphor-based understanding can translate into practical abilities in working with AI systems, and the cross-cultural implications of using metaphors to teach in diverse contexts.

**AI usage disclaimer**

This study used GenAI tools for revision and editorial purposes throughout the production of the manuscript. Models used were ChatGPT (GPT-4o) and Claude 3.5 Sonnet. The authors reviewed, edited and take responsibility for all outputs of the tools used in this study.